\documentclass[12pt,preprint]{aastex}

\shorttitle{MASSES OF THE STARS IN SS CYGNI}
\shortauthors{BITNER, ROBINSON, \& BEHR}

\begin{document}

\title{The Masses and Evolutionary State of the Stars 
in the Dwarf Nova SS Cygni\footnote{Based on observations obtained 
with the Hobby-Eberly Telescope, which is a joint project of the 
University of Texas at Austin, the Pennsylvania State 
University, Stanford University, Ludwig-Maximilians-Universit\"at M\"unchen, 
and Georg-August-Universit\"at G\"ottingen.}}

\author{Martin A. Bitner and Edward L. Robinson}
\affil{Department of Astronomy, 
       University of Texas at Austin, 
       1 University Station C1400, 
       Austin, TX, 78712}
\email{mbitner@astro.as.utexas.edu}
\and
\author{Bradford B. Behr}
\affil{Code 7215, Naval Research Laboratory, Washington, DC 20375}

\begin{abstract}
The dwarf nova SS~Cygni is a close binary star consisting
of a K star transferring mass to a white dwarf by way of
an accretion disk.
We have obtained new spectroscopic observations of SS~Cyg.
Fits of synthetic spectra for  
Roche-lobe-filling stars to the absorption-line spectrum
of the K star yield the amplitude of the K star's 
radial velocity curve and the mass ratio: 
$K_{K} = 162.5 \pm 1.0\ \textrm{km s$^{-1}$}$
and $q= M_{K} /M_{wd} = 0.685 \pm 0.015$.
The fits also show that the accretion disk and white dwarf contribute 
a fraction $f = 0.535 \pm 0.075$ of the total flux at 5500~\AA.
Taking the weighted average of our results with previously
published results obtained using similar techniques, we find
$\langle \textrm{K}_{K} \rangle = 163.7 \pm 0.7\ \textrm{km s$^{-1}$}$
and $\langle q \rangle = 0.683 \pm 0.012$.
The orbital light curve of SS~Cyg shows an ellipsoidal 
variation diluted by light from the disk and white dwarf.
From an analysis of the ellipsoidal variations we limit 
the orbital inclination to the range $45^\circ \le i \le 56^\circ$.
The derived masses of the K star and white dwarf are
$M_K = 0.55 \pm 0.13\ M_\odot$ and
$M_{wd} = 0.81 \pm 0.19\ M_\odot$, where
the uncertainties are dominated by systematic errors
in the orbital inclination.
The K star in SS~Cyg is 10\% to 50\% larger than an unevolved 
star with the same mass and thus does not follow the 
mass-radius relation for Zero-Age Main-Sequence stars; nor 
does it follow the ZAMS mass/spectral-type relation.
Its mass and spectral type are, however, consistent with 
models in which the core hydrogen has been significantly
depleted.

\end{abstract}


\keywords{binaries: close, stars: fundamental parameters, individual: SS Cygni}

%++++++++++++++++++++++++++++++++++++++++++++++++++++++++++
%  INTRODUCTION
%++++++++++++++++++++++++++++++++++++++++++++++++++++++++++

\section{INTRODUCTION}

As the brightest dwarf nova and one of the brightest
cataclysmic variables of any kind, SS~Cygni
has been extensively observed and its
properties are thought to be well established.
Like all cataclysmic variables, it is a
binary star consisting of a cool, relatively normal 
star and a white dwarf \citep{warner95}.
Indeed, it was one of the first cataclysmic variables
shown to be a binary star -- by \citet{joy56}, who
found the orbital period to be $P_{orb} = 0.27624$~d,
later revised to 0.27513~d by \citet{stover80} and
\citet{cowley80}.
The spectral type of the cool star is K4-5~V
\citep{smith98,beuermann00}.
The K star fills its Roche lobe and is transferring mass 
through the inner Lagrangian point to an accretion
disk around the white dwarf.
During quiescence the disk has a low
viscosity and accumulates mass, remaining cool and faint. 
When the surface density of the disk reaches a
critical value, the viscosity increases and
mass flows rapidly through the disk to the white dwarf,
releasing gravitational energy.
The disk becomes hot and luminous, causing the 
dwarf nova outburst \citep{lasota01}.
The outbursts recur roughly every 50 days \citep{cannizzo92}.

Standard techniques for measuring the masses of binary
stars do not work for SS~Cyg because only
the spectrum of the K star is visible and the
orbital light curve does not show eclipses.
Previous measurements of the masses of the stars in
SS~Cyg have, therefore, been forced to rely on dubious 
assumptions about the properties 
of the accretion disk or about the structure of the K star
(see section 5).
In this paper we use a technique for measuring masses
that was first developed for the black hole X-ray binaries
and does not require these assumptions.
The technique has three steps (eg, Casares 2006).
First, the amplitude of the K star's radial velocity curve $K_K$
and the orbital period $P$ can be combined to give the mass function
\begin{equation}
   {{M_{wd} \sin^3 i} \over {(1 + q)^2}}
         = {{K_K^3P} \over {2 \pi G}},
        \label{massfunc-equ}
\end{equation}
where $q=M_K/M_{wd}$ is the ratio of the mass of the K star to
the mass of the white dwarf and $i$ is the orbital inclination.
Second, the rotational velocity of the K star can be measured 
from the rotational broadening of its absorption lines.
If the rotation period of the K star is
tidally locked to the orbital period,
the ratio of its projected rotational velocity 
$V_{rot} \sin i $ to $K_K$ is a function of the mass ratio
\begin{equation}
    {{V_{rot} \sin i} \over { K_K}} = f(q).
\end{equation}
Third, the distorted K star causes ellipsoidal variations
in the orbital light curve. 
The ellipsoidal variations give another relation
between the mass ratio and the orbital inclination.
Light from the accretion disk and white dwarf can distort the 
ellipsoidal variations, so it is also necessary to measure 
the amount of contaminating light from these other sources.
The three relations together yield the dynamical masses 
of the stars and the dimensions of the system.
The derived masses depend only 
weakly on assumed properties of the K star and the accretion disk.

We have obtained new spectroscopic observations of SS~Cyg 
with the Hobby-Eberly Telescope (HET).
We have measured $K_{K}$ and $q$ by fitting the spectrograms
with model spectra calculated using our programs for 
synthesizing the rotationally broadened spectra of cool, 
Roche-lobe-filling stars \citep{bitner06}.
The orbital light curve of SS~Cyg has been measured
by \citet{kjurkchieva99} and by \citet{voloshina00} and
shows a double-humped ellipsoidal variation 
diluted by light from the accretion disk.
We analyze the ellipsoidal variations to place 
limits on the orbital inclination.
Equipped with $K_K$, $q$, and $i$, we derive 
$M_K = 0.55 \pm 0.13\ M_\odot$ and
$M_{wd} = 0.81 \pm 0.19\ M_\odot$.
These masses are significantly smaller than the masses
tabulated in the \citet{ritter98} catalog,
$M_K = 0.704\ M_\odot$ and $M_{wd} = 1.19\ M_\odot$.

Section 2 of this paper describes the spectroscopic
observations and section 3 describes how we measured
$K_K$ and $q$.
In section 4 we describe our analysis of the ellipsoidal 
variations.
Section 5 gives a detailed comparison of our results to earlier 
measurements and discusses the 
reasons for the differences.
In the final section we derive the masses 
of the stars and discuss some implications of the lower masses.

%++++++++++++++++++++++++++++++++++++++++++++++++++
%  OBSERVATIONS
%++++++++++++++++++++++++++++++++++++++++++++++++++

\section{OBSERVATIONS}
The spectrograms of SS~Cyg were acquired with the 
High Resolution Spectrograph (HRS) on 
the Hobby-Eberly Telescope \citep{tull98}.  
We obtained 23 spectrograms in the 30 day interval
from JD~2452084 to JD~2452114 (June and July 2001), 
each covering the wavelength range from 5300~\AA\ to 7000~\AA\ 
at a resolution R = 30,000.
The exposure times were all 600 seconds,
long enough to achieve the necessary signal-to-noise ratio 
without excessive spectral smearing from the changing 
radial velocity of the secondary star as it progressed 
around its orbit.

The heliocentric Julian dates at the midpoints of the
exposures are given in Table~1.
Figure~\ref{EruptionLC-fig} shows 
the eruption light curve of SS~Cyg in 2001 from
the American Association of Variable Star Observers
(http:/www.aavso.org/data/lcg).
The horizontal bar plotted below the light curve 
shows when we obtained our spectrograms.
The figure also shows the eruption
light curve in 1981 when it was observed by \citet{hessman84}.
Both sets of data were obtained when SS~Cyg was in its
quiescent state, but in 2001 the quiescent visual magnitude 
was $m_v = 12.0 - 12.1$, about 0.5 magnitudes fainter than in 1981.

The spectrograms were reduced using standard methods
and standard IRAF packages.
Two instrumental issues are of particular interest 
for the measurements of SS~Cyg.
The HRS has proved to have exceptionally good radial
velocity stability, easily adequate for this 
project \citep{cochran04}.
On the other hand, the HRS is a fiber-fed echelle spectrograph.
Telescope tracking errors, variations in seeing,
atmospheric dispersion, and other instrumental factors 
all limit the spectrophotometric accuracy of HRS spectrograms.
Normalizing the continuum required care, especially 
because the free spectral range of the
orders is rather small ($\sim \! 70$~\AA).

As the optical/near-infrared spectrum of SS~Cyg has been published 
many times \citep{echevarria89, friend90, martinez94, north01,
north02, webb02, harrison00}, we show here only enough of the 
spectrum to demonstrate the rotational broadening of its 
absorption lines.
Figure~\ref{Spectrogram-fig} shows the spectrum of SS~Cyg
near 6120~\AA.  
The spectrum shown is a time-averaged spectrum calculated by 
shifting the 23 individual spectrograms to SS~Cyg's systemic 
velocity and then adding them together.
For comparison the figure also shows a synthetic spectrum for 
a single, non-rotating star with a K4-5~V spectral type 
calculated using the ATLAS9 stellar atmosphere program
and the MOOG spectrum synthesis program.
We show in the next section that the extra width of the 
absorption lines in SS~Cyg is due to a
combination of rotational broadening ($\sim \! 89\ {\rm km\ s}^{-1}$),
orbital smearing during the exposures ($<26\ {\rm km\ s}^{-1}$),
and instrumental resolution of the spectrograph
($8.2\ {\rm km\ s}^{-1}$).
%++++++++++++++++++++++++++++++++++++++++++++++++++
%  K and q
%++++++++++++++++++++++++++++++++++++++++++++++++++

\section{ANALYSIS OF THE SS~CYGNI SPECTRUM}

\noindent
{\it The Synthetic Spectra:}
We measured $K_K$, $q$, and the fraction of the flux contributed
by the accretion disk and white dwarf by fitting the HET spectroscopy
with synthetic spectra for cool, Roche-lobe-filling stars
in close binary systems.
We computed the synthetic spectra with our LinBrod program,
which computes the spectra by summing wavelength-dependent,
velocity-shifted, specific intensities over the distorted 
surface of the lobe-filling star \citep{bitner06}.
The wavelength-dependent specific intensities are 
calculated using the ATLAS9 stellar atmosphere program 
and a modified version of the MOOG spectrum synthesis program
\citep{sneden73,kurucz93}.

The synthetic spectra for the K star depend strongly on the 
orbital period, the orbital phase, the mass ratio, 
and the amplitude of its radial velocity curve;
less strongly on the star's metallicity and temperature;
and only weakly on the orbital inclination, the 
gravity darkening exponent, and other input 
parameters to ATLAS9 and MOOG.
As our primary goal was to determine $K_K$ and $q$, we 
fixed the orbital period at its known value, we assumed
the K star has a solar metallicity, and we adopted 
standard values for the gravity darkening exponent and
the input parameters to ATLAS9 and MOOG.
In preparation for the fits,
we created large grids of synthetic spectra, two dimensions
of the grid covering the possible ranges of $K_K$ and $q$,
and a third dimension covering 100 orbital phases.
Separate grids were calculated for different values of the
orbital inclination and 
effective temperature of the lobe-filling star.
While we did not fit the temperature and
inclination, the separate grids were used to determine the
sensitivity of $K_K$ and $q$ to those parameters and
to measure the dilution of the K star spectrum by
flux from the accretion disk.
To speed the calculation of Doppler shifts the
synthetic spectra were placed on a logarithmic wavelength 
scale with a spacing of 1~km~s$^{-1}$.

The synthetic spectra were convolved with the HRS 
instrumental profile and with an orbital smearing profile.
We approximated the instrumental profile with a trapezoid.
The full width at half maximum (FWHM) of the trapezoid
was set equal to 8.2~km~s$^{-1}$, the
mean FWHM of a dozen isolated, narrow 
lines in spectrograms of the Th-Ar wavelength calibration lamp.
We used a rectangle function to model orbital smearing.
The width of the rectangle at each orbital phase 
was determined from
$\Delta V = K_{K} (2 \pi \Delta t/ P) \cos 2\pi \phi$,
where $\Delta t$ is the length of the exposure and
$\phi$ is the orbital phase.
The zero point in phase is the time of 
superior conjunction of the white dwarf.
For $K_{K} \approx 162\ \textrm{km s}^{-1}$ and 
a 10 minute exposure time the maximum orbital smearing is
$\sim 26$~km~s$^{-1}$.

\noindent
{\it Determination of $K_K$ and $q$:}
The synthetic spectra were fitted to the observed spectrograms
by weighted least squares.
As the spectrograms are greatly oversampled in wavelength, 
we were able to assign a weight to each wavelength from the
internal scatter of the spectrograms.
We fitted only those wavelength regions with several 
strong absorption features.
There were seven of these regions between $5350$~\AA\ 
and $6470$~\AA, each $25$~\AA\ to $57$~\AA\ wide and each
mostly within the central 2/3 of a spectral order. 
As the regions are narrow, we normalized the
continua of both the synthetic and the observed spectra 
to 1.0 by fitting straight lines to the 
continua, a different straight line for each region.

We fitted all the wavelength regions of all the 
spectrograms simultaneously, minimizing the reduced 
$\chi^2$ of the fit to find the best values of $K_K$ 
and $q$, and using the behavior of $\chi^2$ in the 
neighborhood of the minimum to estimate the internal errors.
There are three additional parameters in these fits:
the flux contributed by
the disk and white dwarf, the systemic velocity 
$\gamma$, and the orbital phase offset $\Delta \phi$, 
where the $\Delta \phi$ is measured with respect to 
the \citet{hessman84} ephemeris: 
$T_{0}=$~HJD~2,444,841.86899 and $P=0.27512973$~d.
The flux from the disk and white dwarf was allowed to
run free with a different, independent value for each
of the seven wavelength regions.
As the spectrograms cover the orbit of SS~Cyg fairly
uniformly, we decided to determine $\gamma$ and 
$\Delta \phi$ separately from $K_K$ and $q$.
Initial estimates for $\gamma$ and $\Delta \phi$
were made from fits based on 
initial estimates of $K_K$ and $q$ from 
\citet{beuermann00} and \citet{hessman84}.
Holding these initial estimates for $\gamma$ and $\Delta \phi$ 
fixed, we then re-fitted the spectra to obtain improved values
for $K_K$, $q$, and the diluting flux from the accretion
disk.
With these improved values for $K_K$ and $q$, we looped back
and redetermined $\gamma$ and $\Delta \phi$.
The redetermined values of $\gamma$ and $\Delta \phi$
did not differ significantly
from the initial estimates, and so we were done.

The results are $K_{K} = 162.5\ \pm 1.0\ \textrm{km s$^{-1}$}$,
$q = 0.685 \pm 0.015$, $\Delta \phi = 0.0036 \pm 0.0010$, and 
$\gamma = -16.1\ \textrm{km s$^{-1}$}$.
Figure~\ref{KqChi-fig} shows the contour plot of $\chi^2$ 
as a function of $K_K$ and $q$.
The sign of $\Delta \phi$ is in the sense that
the observed radial velocities occurred a few minutes
later than predicted by the \citet{hessman84} ephemeris.
We do not attach a standard deviation to our measurement
of $\gamma$ because we did not observe radial velocity standards,
leaving the zero point of our radial velocity system uncertain.
The $\gamma$ velocity is, however, consistent with the
heliocentric velocity found by \citet{north02}, 
$\gamma = -13.1 \pm 2.9\ \textrm{km s$^{-1}$}$.

To investigate the dependence of $K_K$ and $q$ on the
effective temperature used to calculate the synthetic 
spectra of the K star, we derived $K_K$ and $q$ 
for two values of $T_{eff}$, 4560~K and 5000~K, the former 
corresponding to a K4~V spectral type and the latter to 
K2~V on the \citet{tokunaga00} temperature scale.
We note in passing that the local effective temperature 
varies considerably from point to point over the surface 
of the star.
The temperatures we quote are the flux-weighted mean
effective temperatures.
The dependence on temperature is weak.
The change of $T_{eff}$ from 4560~K to 5000~K decreased
$K_K$ by only $0.9\ \textrm{km s$^{-1}$}$ and increased $q$
by $0.005$.
The values of $K_K$ and $q$ reported throughout this paper
correspond to 4560~K.

The absorption-line profiles depend on the orbital
inclination, so the derived values of
$K_K$ and $q$ also depend on inclination.
To investigate the strength of this dependence, 
we re-derived $K_K$ and $q$ for several values of $i$.
The dependence is weak.
Changing $i$ from $40^\circ$ to
$50^\circ$ changed $K_K$ by less than $0.5\ \textrm{km s$^{-1}$}$
and $q$ by about $\sim\! 0.001$.
Since the absorption-line profiles depend on the orbital
inclination, they can be used to determine the 
inclination, at least in principle \citep{shahbaz98}.
In practice the dependence is so weak that fits to the 
spectra do not give reliable inclinations \citep{bitner06}.
Indeed, for SS~Cyg our fits would suggest an inclination 
near or lower than $30^\circ$, an impossibly low value
since it would imply a white dwarf mass much greater than
the Chandrasekhar limit.
As a result, we do not determine the inclination this way.

\noindent
{\it Determination of the Flux from the Accretion Disk and White Dwarf:}
The effect of extra flux from the accretion disk and white dwarf is
to dilute the absorption lines in the spectrum of the K star.
Since the depth of the K star's absorption lines depends on
its spectral type, the measured amount of flux from the 
disk and white dwarf depends on the spectral type and
temperature adopted for the star.
The dependence is strong.
For $T_{eff} = 4560\ \textrm{K}$ 
the fits yielded $f = 0.49$, where $f$ is the fraction
of the flux at 5500~\AA\ coming from the disk and white dwarf;
and for $T_{eff} = 5000\ \textrm{K}$ the fits yielded $f = 0.260$.
To measure $f$ we must, therefore, have a more precise value for the 
temperature of the K star than we have needed up to now.
We obtain this temperature in two steps:
First we determine the precise spectral type of the K
star, then we assign a temperature to that spectral
type.

With the exception of \citet{joy56}, who classified the
late type star as dG5, and of \citet{martinez94}, who
classified it as K2-K3, there is a broad consensus that
the late-type star has a K4~V or K5~V spectral type
\citep{stover80,cowley80,walker81,bailey81,wade82, friend90,
harrison00,webb02,north02}.
The spectral type does not depend on the method nor
the wavelength region used for the measurement.
Neither spectral type is preferred over the other.
Thus, in their reviews of cataclysmic variable secondary stars
\citet{smith98} assigned a spectral type of K5~V to the
secondary of SS~Cyg while
\citet{beuermann00} assigned a spectral type of K4~V.
We adopt K4.5~V with a half subclass uncertainty. 
The surface of the K star is likely to have regions
with much lower temperatures (starspots) \citep{webb02}.
Their effect on the spectral type is assumed
to be negligible.

From Table~7.6 in \citet{tokunaga00} the temperature
corresponding to spectral type K4.5~V is 4450~K, while from
Table~15.7 in \citet{drilling00}, the temperature is
4480~K.
Allowing for a half spectral-subclass uncertainty and a possible
systematic error of 100~K due to starspots, heating by 
irradiation, or an incorrect temperature scale, we take
$T_{eff} = 4470 \pm 140$~K as the best
estimate of the temperature of the K star.

As noted above, for $T_{eff} = 4560$~K we find
$f = 0.49$, and for $T_{eff} = 5000$~K we find $f = 0.260$.
Extrapolating somewhat, we find $f = 0.535 \pm 0.075$ for 
$T_{eff} = 4470 \pm 140$~K.
We note that errors in assumed metallicty can affect the line 
depths in our model LinBrod spectra and thus the derived flux 
contribution from the accretion disk and white dwarf.
To quantify the magnitude of the effect, we compared results for 
LinBrod model spectra of solar metallicty and [M/H] = -0.5.
This is a much larger metallicity error than we expect is reasonable.
Even so, the derived flux contribution from the white dwarf and 
accretion disk only drops by 20-25\% for the lower metallicity spectra.
We measured the disk flux in each of the seven wavelength
regions separately.
While the results are rather noisy, the flux contributed
by the disk is roughly 20\% less at 6400~\AA\ than
at 5500~\AA.
Other spectroscopic measurements of the 
disk fraction are listed in Table~2 for comparison.
With the possible exception of the measurement by
\citet{north02} there is surprisingly good agreement
among them.

\noindent
{\it An External Check on the Results:}
As a final external check on our results one of us (BBB)
re-reduced the spectrograms of SS~Cyg with
the Figaro package and re-analyzed them with the LINFOR
spectral synthesis package
(Baschek et al.\ 1966; Lemke 1997, private communication), 
both of which are genetically 
unrelated to the IRAF and LinBrod packages.
Using LINFOR we generated a  model spectrum for the 
secondary, including a spherically-symmetric rotation 
profile with $v \sin i = 100$~km/s, and cross-correlated this 
model with each observed spectrum to get initial radial
velocities. 
We then adjusted the model rotational velocity and the abundances 
of the major line-forming atomic species to minimize the 
residuals between the model and the observations. 
Finally, we used the model to identify regions of continuum 
and pseudo-continuum, and adjusted the adopted continuum levels 
for the observed spectra accordingly. 
The entire sequence was repeated until the results converged.
We list the resulting radial velocities in Table~1.
The rotational velocity varied from about 
84 to $94\ \textrm{km\ s}^{-1}$, depending on orbital phase, 
with a mean near $V_{rot} \sin i = 89\ \textrm{km\ s}^{-1}$.
This method does not yield a reliable mass 
ratio because of the spherical approximation,
and it does not yield a measure of the dilution because
the dilution is mimicked by a reduction of metallicity;
but the radial velocities should be unbiased \citep{bitner06}.
A fit to the velocities yielded 
$K_K = 162.8\ \textrm{km s$^{-1}$}$ in excellent agreement
with the LinBrod results.

%++++++++++++++++++++++++++++++++++++++++++++++++++
%  LIGHT CURVE
%++++++++++++++++++++++++++++++++++++++++++++++++++

\section{LIGHT CURVE ANALYSIS AND THE ORBITAL INCLINATION}

\noindent
{\it The Source of the Orbital Variations:}
\citet{kjurkchieva99} and \citet{voloshina00}
have independently published multicolor mean orbital 
light curves for SS~Cyg and the two sets of light curves
agree to $\sim \!0.01$~mag.
We use the \citet{voloshina00} data
for our analysis because they give the variances 
of the individual points in the light curves.
The mean V-band light curve from 
\citet{voloshina00}, reproduced in Figure~\ref{LCfit-fig},
shows a distinctive double-humped modulation with 
minima near orbital phases $\phi = 0.0$ and 0.5,
and unequal maxima near $\phi = 0.25$ and 0.75,
where phase zero is again defined to be superior conjunction
of the white dwarf.
We attribute the double humped modulation primarily 
to ellipsoidal variations of the K star.  
The best justification for this interpretation is 
the quantitative agreement we find between the observed 
light curves and synthetic light curves based on 
ellipsoidal variations.
The double humps have occasionally
been attributed to spots on the white dwarf, caused perhaps
by accretion onto magnetic poles \citep{kjurkchieva99}.
Since the humps remain at the same orbital phases, this
interpretation would require the that rotation period of the
white dwarf be identical to the orbital period.
There is no evidence that SS~Cyg has 
a magnetic field strong enough to lock on to the K star 
and enforce synchronism (ie, it is not an AM Her star) and,
on the contrary, there is evidence that the rotation period
of the white dwarf is just a few seconds \citep{mauche04}.
Spots on the white dwarf cannot then account for the
double humped modulation.

A pure ellipsoidal variation produces two 
symmetric humps in a light curve, but in SS~Cyg the hump 
at $\phi = 0.75$ has a larger amplitude than the hump at 
$\phi = 0.25$.
The multicolor light curves published
\citet{kjurkchieva99} and \citet{voloshina00}
show that difference between the two humps becomes 
progressively greater at shorter wavelengths,
demonstrating that the extra flux in the larger hump is bluer
than the flux in the pure ellipsoidal variations.
The disks in cataclysmic variables generally have a hot spot
on their edge where they are hit by a stream of gas transferred
from the lobe-filling star, and the hot spot often produces
a single hump in the orbital light curve peaking at 
phases between 0.7 and 0.9 \citep{warner95}.
Since the extra blue flux in SS~Cyg is greatest near 
phase 0.75, we attribute the extra blue flux to a hot spot
on the disk.
Our interpretation of the light curve is, thus, similar 
to that of \citet{voloshina00}.

\noindent
{\it A Model for the Light Curve:}
Our goal is to measure the orbital inclination of
SS~Cyg from the amplitude of the ellipsoidal variations.
The ellipsoidal variations are, however, diluted by 
unmodulated light from the accretion disk.
This complicates the task because the ellipsoidal 
variations of binaries with low inclinations and 
low dilution are nearly identical to those of binaries 
with higher inclinations and greater dilution.
\citet{voloshina00} attempt to break the degeneracy 
between inclination and dilution by 
fitting multicolor light curves.
For this method to work, the spectral energy 
distributions of all sources of flux must be known.
They assumed that all the sources emit black body radiation.
This is a poor assumption for the optical flux from 
the quiescent accretion disks of dwarf novae 
\citep{vrielmann02,saito06}
and is not really valid even for the flux from the 
two stars, making the orbital inclination 
determined from the multicolor fits unreliable.

The degeneracy between the orbital inclination and
the disk flux can also be broken if the disk flux
can be independently measured.
The disk flux then becomes an input parameter, not a
fitted parameter, and we can dispense with theoretical
models for the disk flux and spectral energy distribution.
From the fits to the spectrum described in section 3, 
we do know the V-band flux from the disk, so
this is the method we will use to break the degeneracy.

The computer program we use to calculate light curves 
for SS~Cyg is an updated and improved version of the 
program we have used to calculate the light curves of black 
hole X-ray binaries \citep{ioannou04}.
Although the program accepts many parameters and can synthesize
light curves for complicated accretion geometries, just three 
components are needed to fit the light curve
of SS~Cyg:
\begin{itemize}
\item
The Ellipsoidal Variations:
We use Roche geometry to calculate the shape and local effective
gravity of the lobe-filling K star.
The local effective temperature $T_{eff}(local)$ is calculated from
the gravity darkening law 
$T_{eff}(local) \propto T_{eff} |g|^\beta$, where
$g$ is the local gravity, and $\beta$ is the gravity-darkening
exponent.
According to \citet{claret00a} $\beta \approx 0.10$ for
main-sequence stars with temperatures near 4500~K.
The amplitude of the ellipsoidal variations depends only
weakly on $\beta$.
The angle-averaged fluxes in specific filters are calculated 
by integrating Kurucz model spectra over the 
filter response functions.
We adopt the \citet{claret00b} limb darkening law
to convert the fluxes to the local specific intensities,
which are then integrated over the surface of the star
to produce the observed brightness.

\item
Constant Flux from the Disk:
The main body of the accretion disk is  
uneclipsed and just adds constant extra flux to the light curve.
Any contribution from the white dwarf is subsumed
in the disk contribution.

\item
Variable Flux from a Bright Spot on the Disk:
The disk has a bright spot painted on its outer edge.
The spot is visible only when on the side of the accretion
disk facing the observer and its brightness is
modulated by geometric foreshortening.
The properties of the spot are specified by
its position and extent along the disk edge and by
its brightness.
\end{itemize}

\noindent
{\it The Fits to the Light Curve:}
Although \citet{voloshina00} published light curves in the 
U, B, and V bands, we fit only the V-band light curve 
because the distorting effect of the bright spot is least 
in that filter and because the V band is the only one 
for which our spectroscopy gives
a measurement of the diluting flux from the disk.
For all fits we use $K_{K} = 162\ \textrm{km s}^{-1}$ and
$q = 0.69$.
The deduced orbital inclinations are essentially independent
of $K_{K}$, and
changing the mass ratio by $\Delta q = 0.01$
changes the deduced orbital inclinations by just $0.5^\circ$.
The errors introduced by uncertainties in $K_{K}$ and $q$
are, therefore, small compared to other sources of error.
Since we are fitting just the V-band light curve, the
deduced inclinations are almost independent of the adopted 
effective temperature, with just a second-order dependence
through the limb-darkening and the gravity darkening coefficient.
For convenience internal to the computer program we used 
$T_{eff} = 4600$~K, not 4470~K.
Numerical tests showed that the ellipsoidal light curves
and the derived inclinations are not significantly changed
by the difference in temperature.

We proceeded by choosing a specific value for the fraction 
of the V-band flux coming from the disk 
(at $\phi = 0.25$) and 
then fitting the synthetic light curve to the data by
adjusting the remaining parameters to minimize the 
$\chi^2$ of the fit.
The four physically-significant parameters are the 
orbital inclination, the location of the spot on the 
edge of the disk, the extent of the spot 
along the edge, and the total V-band flux from the spot.
There are also two nuisance parameters: the
offset in orbital phase between the synthetic and
observed light curves, and an overall scale factor.
We repeated the fit for many different values of the
fraction of the flux coming from the disk, 
producing a table of orbital inclinations and 
standard deviations as a function of the disk flux.

Figure~\ref{LCfit-fig} shows a synthetic light curve 
over-plotted on the observed light curve.
The synthetic light curve corresponds to an orbital
inclination of $52^\circ$ with 56\% of the light 
coming from the accretion disk and a phase offset
of $\Delta \phi = 0.01$.
The amplitude of the ellipsoidal variations -- and therefore
the deduced orbital inclination -- is determined almost
entirely by the amplitude of the hump at phase 0.25,
whereas the properties of the spot are 
determined almost entirely by the hump at phase 0.75.
As a result, the orbital inclinations depend only weakly 
on the spot model.
Figure~\ref{inclination-fig} shows the deduced orbital 
inclination as a function of the fraction
of the V band flux coming from the accretion disk.
The best fit inclinations are shown by the solid line 
extending from the lower-left to the upper right in the diagram.
The dashed lines above and below the solid line show the
one standard deviation confidence limits for the fits.

\noindent
{\it The Orbital Inclination:}
In the previous section 
we found that the fraction of the V band flux
coming from the disk was $f = 0.535 \pm 0.075$.
The vertical solid line in Figure~\ref{inclination-fig}
corresponds to $f = 0.535$ and the two vertical dashed lines 
to $f = 0.46$ and $f = 0.61$.
The allowed values of the disk flux and orbital inclination
lie in the region enclosed by the four dashed lines in the figure.
Rounding the artificially sharp corners of the region
somewhat, we find that the orbital inclination of SS~Cyg 
lies in the interval $45^\circ \le i \le 56^\circ$.
The range of possible inclinations is rather wide.
The width is dominated by the uncertainty in $f$, which traces
back to the uncertainty in the temperature of the K star determined
from its spectral type; other sources of error
are minor contributors.

%++++++++++++++++++++++++++++++++++++++++++++++++++
%  COMPARISON TO PREVIOUS RESULTS
%++++++++++++++++++++++++++++++++++++++++++++++++++

\section{COMPARISON TO PREVIOUS RESULTS}

As the literature on SS~Cyg is replete with discrepant
measurements of $K_K$, $q$, $i$ and the masses of its components,
and as our measurements lead to yet another set of masses,
it is appropriate to give a critical review of the literature,
comparing our results to 
the previous measurements and analyzing the reasons for the differences.

{\it Radial Velocity Curve of the K Star:}
Table~3 lists all the published measurements of $K_K$.
The amplitudes fall into three groups.
The first consists of early measurements
yielding amplitudes near $120\ \textrm{km s$^{-1}$}$.
In this group the spectrograms were recorded on photographic 
plates and velocities were measured using traditional techniques.
Beginning with \citet{stover80} the spectrograms were obtained 
with digital detectors and the velocities were usually measured
by cross-correlating or directly fitting a template spectrum
to the spectrograms.
Initially controversial, Stover's measurement was the first
of a group made between 1980 and 1990 yielding 
amplitudes near $155\ \textrm{km s$^{-1}$}$.
There are two main reasons for the difference between these
and the earlier results.
First, the absorption lines from the K star
are broadened by rotation and veiled by continuum emission
from the accretion disk, making
line positions difficult to measure on photographic plates.
Second, heating of the K star by irradiation from the accretion
disk during dwarf nova outbursts distorts the
radial velocity curve, changing the measured
amplitude \citep{robinson86}.
While the distortion has not been fully modeled theoretically,
the presence of the distortion is amply demonstrated by observations
\citep{martin89,echevarria89}.
Observers have since become careful to measure the radial 
velocity curve of the K star during quiescence, although
\citet{robinson86} suggested that there might be enough 
irradiation to bias the measured value of $K_K$ significantly 
even during quiescence.

The radial velocity amplitudes measured by \citet{martinez94} and
\citet{north02}, $162.5 \pm 3$ and $165 \pm 1\ \textrm{km s$^{-1}$}$ 
respectively, and now $162.5\ \pm 1.0\ \textrm{km s$^{-1}$}$
by us,
comprise the third group.\footnote{\citet{martinez94} 
measured $162.5 \pm 3\ \textrm{km s$^{-1}$}$ from their 
own data but then adopted $158 \pm 3\ \textrm{km s$^{-1}$}$ 
from \citet{robinson86} in the rest of their paper.}
While the difference between these amplitudes and the 
$\sim \! 155\ \textrm{km s$^{-1}$}$ amplitudes of the second
group is small, it is statistically significant.
We attribute the difference to a 
real change in the SS~Cyg system.
From \citet{cannizzo92} the mean quiescent visual magnitude
of SS~Cyg between 1896 and 1992 was $m_v = 11.8 - 11.9$.
The quiescent magnitude fluctuated by a few tenths of a magnitude
about the mean and appeared to show a $\sim 0.1$ magnitude 
secular decrease over the 100 year interval.
\citet{hessman84} observed SS~Cyg when its quiescent magnitude 
was $m_v = 11.5 - 11.6$, about 0.3 magnitude brighter than the
historical average (see Figure~\ref{EruptionLC-fig}).
AAVSO light curves (http:/www.aavso.org/data/lcg) show
that SS~Cyg was at $m_v = 11.9 - 12.0$ when observed
by \citet{martinez94}, at $m_v \approx 12.0$ when observed
by \citet{north02}, and from Figure~\ref{EruptionLC-fig} 
at $m_v = 12.0 - 12.1$ when observed by us.
Succinctly, the recent observations were obtained when
SS~Cyg was fainter than its historical average and much
fainter than when observed by \citet{hessman84}.
Following \citet{cannizzo92} and \citet{warner95}, we 
attribute a lower quiescent brightness to a lower 
accretion luminosity, which implies lower
irradiation of the K star and less distortion of its
radial velocity curve.
As a result, the values of $K_K$ in the third group should
be a better representation of the true motion of the
K star.

{\it The Mass Ratio:}
As the white dwarf in SS~Cyg is a minor contributor to the
flux at optical and infrared wavelengths, its radial velocity
curve cannot be directly measured and the mass ratio must be 
deduced by other means.
Lacking a better alternative it has sometimes been
assumed that the 
wings of the emission lines, which come from high-velocity gas 
near the white dwarf, can serve as a proxy for the white
dwarf and track its orbital motion.
Measurements of the amplitude $K_{em}$ of the emission-line 
radial velocity curve of SS~Cyg are shown in Table~3.
The weighted average of the amplitudes measured with
electronic detectors is
K$_{em} \approx 92\ \textrm{km s$^{-1}$}$.
Taking $\textrm{K}_{K} \approx \textrm{K}_{abs} 
\approx 152 - 165\ \textrm{km s$^{-1}$}$, one finds
$q = M_{K}/M_{wd} = 0.56 - 0.60$, in accord with the
value tabulated by \citet{ritter98}.
Although one of us has previously argued that emission lines 
can yield good qualitative estimates of the mass ratios
\citep{robinson92}, more recent work has demonstrated
that emission-line velocities can yield biased or simply wrong
mass ratios and should not be used for accurate 
work [eg, \citet{welsh93}].

We have measured the mass ratio from the rotational 
broadening of the absorption lines in the spectrum 
of the K star.
This technique has been widely used to measure the
masses of black holes in black hole binary stars
and, if applied with care, gives reliable mass ratios
\citep{casares06}.
\citet{cowley80} used line broadening as a consistency 
check on other ways to measure the mass ratio 
of SS~Cyg but \citet{martinez94} were the first to 
use line broadening as the fundamental way
to measure the mass ratio, finding $q = 0.62 \pm 0.05$.
Their measurement was, however, biased by the 
adoption of a \citet{gray92} rotational broadening profile, 
which is
only appropriate for spherical stars, and by the use of
approximate formulae for converting from rotational velocity
to mass ratio.
\citet{north02} also measured the mass ratio from the
rotational broadening of the K star's spectrum.
Their method, outlined in \citet{north00}, is similar
to ours and like ours should be free of bias. 
Their result, $q = 0.68 \pm 0.02$, is nearly identical 
to ours.

{\it The Orbital Inclination:}
SS~Cyg does not eclipse, so the orbital inclination must 
be measured indirectly.
One way has been to assume the K star has some known
property and then work backwards to the orbital inclination.
If one assumes that the K star follows the mass/spectral type relation
for main-sequence stars, its mass is
$0.7 - 0.8\ M_\odot$ and the inferred orbital inclination
is $38^\circ - 40^\circ$ \citep{cowley80,walker81}.
It is more common to assume that the K star follows the
main-sequence mass/radius relation.
There are many variants of this relation but all lead to 
orbital inclinations in the range $35^\circ - 45^\circ$
\citep{kraft62,kiplinger79,stover80,hessman84,hessman88,friend90}.
The properties of lobe-filling stars can, however, depart 
greatly from main-sequence relations because of 
nuclear evolution before the onset of mass transfer and the loss of
thermal equilibrium while mass is being transferred.
The theoretical calculations by \citet{kolb01} show that 
lobe-filling stars with K4 -- K5 spectral types can have
masses anywhere in the range $0.42 - 0.80\ M_\odot$.
\citet{baraffe00} show that the radii of stars that have used up
some of the hydrogen in their cores and are transferring mass
can be larger than the radii of normal main-sequence
stars with the same mass by 50\% or more.
The observational data collected by \citet{smith98} and 
\citet{beuermann00} support these calculations.

There have been several attempts to estimate the orbital 
inclination from properties of the accretion disk.
Comparing the profile of the H$\beta$ emission line in SS~Cyg
to theoretical profiles for a thin disk, 
\citet{cowley80} derived an inclination between $30 - 40^\circ$.
If the inner radius of the accretion disk is only slightly greater
than the radius of the white dwarf and if the white dwarf obeys 
the usual white dwarf mass/radius relation, one can use the 
inner radius to deduce the mass of the white dwarf and again 
work backwards to the orbital inclination.
\citet{cowley80} estimated the radius from the maximum velocity 
extent of the wings of the Balmer emission lines,
finding $i \approx 46^\circ$.
\citet{giovannelli83} assumed that the periods of 
dwarf nova oscillations (DNOs) are equal to the orbital 
periods of gas at the inner edge of the accretion disk.
At that time the DNOs of SS~Cyg had been observed to range from
7.3~s to 10.9~s, from which \citet{giovannelli83}
derived $i = 40^{\circ +1^\circ}_{\phantom{\circ}-2^\circ}$.
Doppler tomograms demonstrate, however, that the velocity
fields of accretion disks, including the accretion disk
of SS~Cyg, are complicated \citep{robinson93,north01}, and the
velocities are likely to be especially complicated near the 
white dwarf 
where magnetic fields may play a role; and the periods of 
DNOs are only indirectly related to Keplerian periods 
\citep{warner04,kluzniak05},
all of which precludes giving much weight to inclinations
derived from properties of the accretion disk.

The analysis of orbital light curves generally yield
reliable inclinations.
This is clearly true for eclipsing binaries but extensive 
observations of the black hole binaries have shown that 
this is also true for orbital inclinations derived from 
the ellipsoidal variations of Roche-lobe-filling stars
\citep{charles06}.
Ellipsoidal variations are produced primarily by the
geometric distortions of the star, which, since the star fills
its Roche lobe, are a function only of the mass ratio.
Large amplitude ellipsoidal variations imply a large
inclination, low amplitude a low inclination.
The rapid flickering in the light curve of
SS~Cyg and its awkwardly-long orbital period vitiated earlier 
attempts to measure its ellipsoidal variations 
(eg, Honey et al.\ 1989).
The variations are, however, present and
have been observed at both ultraviolet and optical wavelengths 
\citep{bartolini85,voloshina86,lombardi87,bruch90,kjurkchieva99,
voloshina00}.
The ellipsoidal variations of SS~Cyg are diluted by
flux from the accretion disk and this dilution must
be included in the models if fits to the light curves
are to give reliable inclinations. 
\citet{voloshina00} dealt with this by assuming all 
components of the SS~Cyg system radiate like black bodies,
finding $i = 51^\circ - 56^\circ$, but their assumption is
questionable, raising concerns about the 
reliability of their result.
We have estimated the disk contribution externally, 
from the dilution of the K star's spectrum.
Consequently we were able to use a far simpler model for 
the light curve, which leads in turn to a more robust 
estimate of the orbital inclination.
Nevertheless, the range of inclinations we find, 
$ 45^\circ \le i \le 56^\circ$, is not incompatible
with the range of inclinations found by \citet{voloshina00}.
The qualitative reason why orbital inclinations derived
from the ellipsoidal variations is high
is that the observed ellipsoidal variations have
a large amplitude even though they are heavily diluted 
by flux from the accretion disk.
Unless the measured orbital light curves incorrectly
represent the true light curve, it is difficult to
escape the higher inclination.

%++++++++++++++++++++++++++++++++++++++++++++++++++
%  DISCUSSION
%++++++++++++++++++++++++++++++++++++++++++++++++++

\section{DISCUSSION}

{\it The Masses of the Stars:}
From fits of synthetic spectra to spectrograms of
SS~Cyg obtained with the HET we have found the amplitude
of the K star's radial velocity curve to be
$K_{K} = 162.5 \pm 1.0\ \textrm{km s$^{-1}$}$ and 
the ratio of the masses of the two stars to be
$q = M_{K} /M_{wd} = 0.685 \pm 0.015$.
From an analysis of the ellipsoidal variations in the orbital
light curve plus an estimate of the dilution of the ellipsoidal
variations by flux from the accretion disk,
we have limited the orbital inclination to the range
$45^\circ \le i \le 56^\circ$.

Following our discussion in section 5, we
see no reason not to accept the measurements of $K_K$
by \citet{martinez94} and \citet{north02} and the
measurement of $q$ by \citet{north02}.
The best estimates of $K_K$ and $q$ are, then, the weighted 
average of their measurements and ours,
$\langle K_K \rangle = 163.7 \pm 0.7\ \textrm{km s$^{-1}$}$
and 
$\langle q \rangle = 0.683 \pm 0.012$.
Adopting these values and taking $45^\circ \le i \le 56^\circ$,
we find masses
$M_K = 0.55 \pm 0.13\ M_\odot$
and
$M_{wd} = 0.81 \pm 0.19\ M_\odot$.
The uncertainties in the masses are dominated by the uncertainty
in the orbital inclination, which is itself dominated by
systematic errors in the temperature of the K star.
As a result, it is best to interpret these values as the ranges 
of possible masses and the midpoints of the ranges,
not the best masses and their standard deviations.
A Monte-Carlo error analysis suggests that lower masses
within the range are somewhat more likely than higher masses.
The masses of the K star and the white dwarf cannot be 
chosen independently from within their ranges as they are
strongly constrained by the mass ratio
$M_{K}/M_{wd} = 0.683 \pm 0.012$.
Kepler's third law now yields the separation
of the centers of mass of the two stars
$a = (1.36 \pm 0.11) \times 10^{11}$~cm, where higher 
values of $a$ correspond to higher masses.

{\it The Evolutionary State of the K Star:}
If the mean radius of a Roche lobe is $R_L$, then
$R_L/a$ is a function only of the mass ratio,
$R_L/a = F(q)$.
Since SS~Cyg is a mass transfer system, the radius of the
K star $R_K$ equals the radius of its Roche lobe and 
can be calculated from $R_K/a = F(q)$.
Adopting the \citet{eggleton83} relation for $F(q)$ and
using our measurement of $a$ and $M_K$ as a function of $i$, we
find $R_K/M_K = 1.07$ at $(i = 45^\circ$,\ $M_K = 0.68)$,
increasing to $R_K/M_K = 1.47$ at $(i = 56^\circ$,\ $M_K = 0.42)$,
where the units of $R$ and $M$ are solar radius and solar mass.

From the theoretical models of \citet{chabrier97} 
the mass-radius relation for zero age main sequence (ZAMS)
stars with masses between $0.4\ M_\odot$ and $0.6\ M_\odot$ is
$R = 0.95 M$, or $R/M = 0.95$,
where here also the units are solar radius and solar mass.
Thus, the K star in SS~Cyg is 10\% to
50\% larger than a single unevolved star with the same mass --
it does not obey the mass-radius relation for unevolved
main-sequence stars.\footnote{If we were to discard the 
orbital inclination determined from the ellipsoidal 
variations and insist that the K star obeys the 
\citet{chabrier97} ZAMS mass-radius relation,
we would find $M_K = 0.81\ M_\odot$,
$M_{wd} = 1.18\ M_\odot$, and $i = 42^\circ$.
These are stringent upper limits on the masses and 
lower limit on the orbital inclination.}
Nor does the K star follow the ZAMS mass/spectral-type 
relation for Roche-lobe-filling stars.
\citet{kolb01} have calculated 
theoretical evolutionary models for the secondary stars 
of mass-transfer binaries.
They find that an unevolved lobe-filling star 
with spectral type K4-5 (their ``standard sequence'')
should have a mass $\sim 0.2\ M_\odot$ greater than the mass
of the K star in SS~Cyg.
Finally, according to \citet{howell01} an unevolved lobe-filling star
in a binary with SS~Cyg's 6.6~hr orbital period 
should have a mass $\sim 0.2\ M_\odot$ greater than 
that of the K star in SS~Cyg.

The \citet{kolb01} models do, however, predict lower masses
for K4-5 stars in which core hydrogen is depleted but 
not exhausted. 
If the mixing length is maintained at $\alpha = 1.0$,
the depletion must be extreme to match SS~Cyg, 
$X_c = 4 \times 10^{-4}$; but if
the mixing length is increased to $\alpha = 1.9$, the
depletion can be more moderate.
We interpret our results, therefore, as showing that the K star 
in SS Cyg has significantly depleted its hydrogen.

\citet{harrison04} showed that the CO absorption bands in the 
infrared spectrum of SS~Cyg are anomalously weak and
\citet{harrison05} found weak CO bands in 9 of 
12 other cataclysmic variables they observed.
They attribute the weak CO bands to a low C abundance
and note that C can be depleted by CNO processing.
While the CNO cycle can indeed deplete C, all that is really 
needed is to process C to N;
the full CNO cycle is not required.
In either case C should not be depleted in 
normal main-sequence stars, so the weak CO bands are evidence
for nuclear evolution.
We take this as supporting evidence that the K star 
in SS~Cyg is not an unevolved main-sequence star.

{\it The Visual Magnitude of SS Cyg:} 
To calculate the expected V magnitude of
SS~Cyg, we begin with the absolute visual
magnitude of a normal K4.5~V star, which
from Table~15.7 of \citet{drilling00} 
is $M_V = 7.19$.
The radius of the K star in SS~Cyg is
10\% to 50\% larger than normal.
Scaled by its surface area the K star's absolute magnitude 
should be greater than normal by 0.21 to 0.88 magnitude, or
$M_{V_K} = 6.64 \pm 0.34$.
The trigonometric parallax of SS~Cyg is 
$6.02\pm0.46$~milliarcseconds, leading to a distance modulus 
$\Delta m = 6.10 \pm 0.16$.
The apparent visual magnitude of the
K star alone should then be $V_K = 12.74 \pm 0.37$.
This must be augmented by the flux from the accretion
disk and white dwarf, which we have measured to be a fraction
$f = 0.535 \pm 0.075$ of the total V-band flux.
The extra flux raises the expected V magnitude of the
entire system to $V = 11.92 \pm 0.40$,
which is consistent with the visual
magnitude $m_v = 12.0 - 12.1$ measured by the
AAVSO at the time of our observations.

\acknowledgements
We thank the American Association of Variable Star Observers
for the data used to make Figure~1, Carlos Allende Prieto 
for help with HET HRS data reduction, Robert A.\ Wittenmyer
for performing an independent fit of a spectroscopic orbit
to the velocities listed in Table~1, William F.\ Welsh
for comments on a draft of this paper and for performing the
Monte-Carlo error analysis, and Chris Sneden for help using and 
modifying the MOOG program.  This research was supported by 
NSF grant AST-0206029.  MAB acknowledges support from NSF grant AST-0607312.  
This research has made use of NASA's Astrophysics Data System (ADS) 
Bibliographic Services.  The Hobby-Eberly Telescope (HET) is a joint 
project of the University of Texas at Austin, the Pennsylvania State 
University, Stanford University, Ludwig-Maximilians-Universit\"at M\"unchen, 
and Georg-August-Universit\"at G\"ottingen.  The HET is named in honor 
of its principal benefactors, William P. Hobby and Robert E. Eberly.

%++++++++++++++++++++++++++++++++++++++++++++++++++++
%  REFERENCES
%++++++++++++++++++++++++++++++++++++++++++++++++++++

\clearpage

\clearpage
\begin{deluxetable}{cccc}
\tablewidth{4.7in}
\tablecaption{Radial Velocities of the Secondary Star in SS~Cyg}
\tablehead{
\colhead{Date} & 
\colhead{Velocity} & 
\colhead{Date} & 
\colhead{Velocity} \\
\colhead{(HJD $-$ 2452000)} & 
\colhead{(km s$^{-1}$)} & 
\colhead{(HJD $-$ 2452000)} & 
\colhead{(km s$^{-1}$)}
}
\startdata
  83.81229  &  136.2  &  104.77608 &  104.7  \\
  86.80263  &   31.0  &  105.76826 &  -34.6  \\
  91.80944  &  150.7  &  106.74746 &  -56.2  \\
  94.80816  &  108.8  &  107.78210 &  143.3  \\
  95.79076  & -170.5  &  109.76639 &   -8.5  \\
  97.76919  &  -95.9  &  110.77827 &  138.2  \\
  98.80362  & -155.5  &  111.73070 & -139.7  \\
  99.78387  &  144.1  &  112.74418 &  122.4  \\
 100.75669  & -174.9  &  113.72400 & -110.2  \\
 101.78850  &  -30.6  &  113.76127 &   18.6  \\
 102.79008  &  121.4  &  113.94938 & -174.8  \\
 103.78679  &  -168.5 &            &         \\
\enddata
\end{deluxetable}

\clearpage
\begin{deluxetable}{ccl}
\tablewidth{5.0in}
\tablecaption{Measurements of the Contribution of Disk Flux to the 
Total Flux from SS~Cygni during Quiescence}
\tablehead{ 
\colhead{Wavelength} & 
\colhead{Fraction of Flux}      & 
                        \\
\colhead{Region} &
\colhead{from the Disk} &
\colhead{Reference}     
} 
\startdata
  near H$\beta$    & $0.50\pm0.03$     & \citet{hessman84} quoted    \\
                   &                   & \qquad in \citet{hessman88} \\
  5500\ \AA        & $0.56 \pm 0.03$   & \citet{wade82} \\
  5500\ \AA        & $0.575 \pm 0.075$ & This work      \\
  6300 - 6500\ \AA & $0.315 \pm 0.004$ & \citet{north02} \\
  6400 - 6700\ \AA & 0.45              & \citet{martinez94} \\
  7000 - 7600\ \AA & 0.50              & \citet{webb02}   \\
\enddata
\end{deluxetable}
\clearpage
\begin{deluxetable}{ccl}
\tablewidth{4.0in}
\tablecaption{Previous Measurements of the Emission- and Absorption-line
 Radial Velocity Curves of SS Cygni}
\tablehead{ 
\colhead{K$_{K}$}     & 
\colhead{K$_{em}$}      & 
                        \\
\colhead{(km s$^{-1}$)} &
\colhead{(km s$^{-1}$)} &
\colhead{Reference}     
} 
\startdata
      115      &     122     & \citet{joy56}        \\
      115      &     122     & \citet{walker68}     \\
       --      & $ 85 \pm 4$ & \citet{kiplinger79}  \\
 $120 \pm 6$   & $118 \pm 8$ & \citet{cowley80}     \\
 $123 \pm 2$   & $107 \pm 2$ & \citet{walker81}     \\
 $153 \pm 2$   & $ 90 \pm 2$ & \citet{stover80}     \\
 $155 \pm 2$   & $ 96 \pm 3$ & \citet{hessman84}    \\
 $152 \pm 2$   & $ 96 \pm 5$ & \citet{echevarria89} \\
 $155 \pm 3$   & $ 92 \pm 2$ & \citet{friend90}     \\
 $162 \pm 3$   &      --     & \citet{martinez94}   \\
 $165 \pm 1$   &      --     & \citet{north02}      \\
 $162.5 \pm 1$ &      --     & This work
\enddata
\end{deluxetable}

%++++++++++++++++++++++++++++++++++++++++++++++++++++
% FIGURE CAPTIONS
%++++++++++++++++++++++++++++++++++++++++++++++++++++

\clearpage

%  FIGURE 1
\figcaption[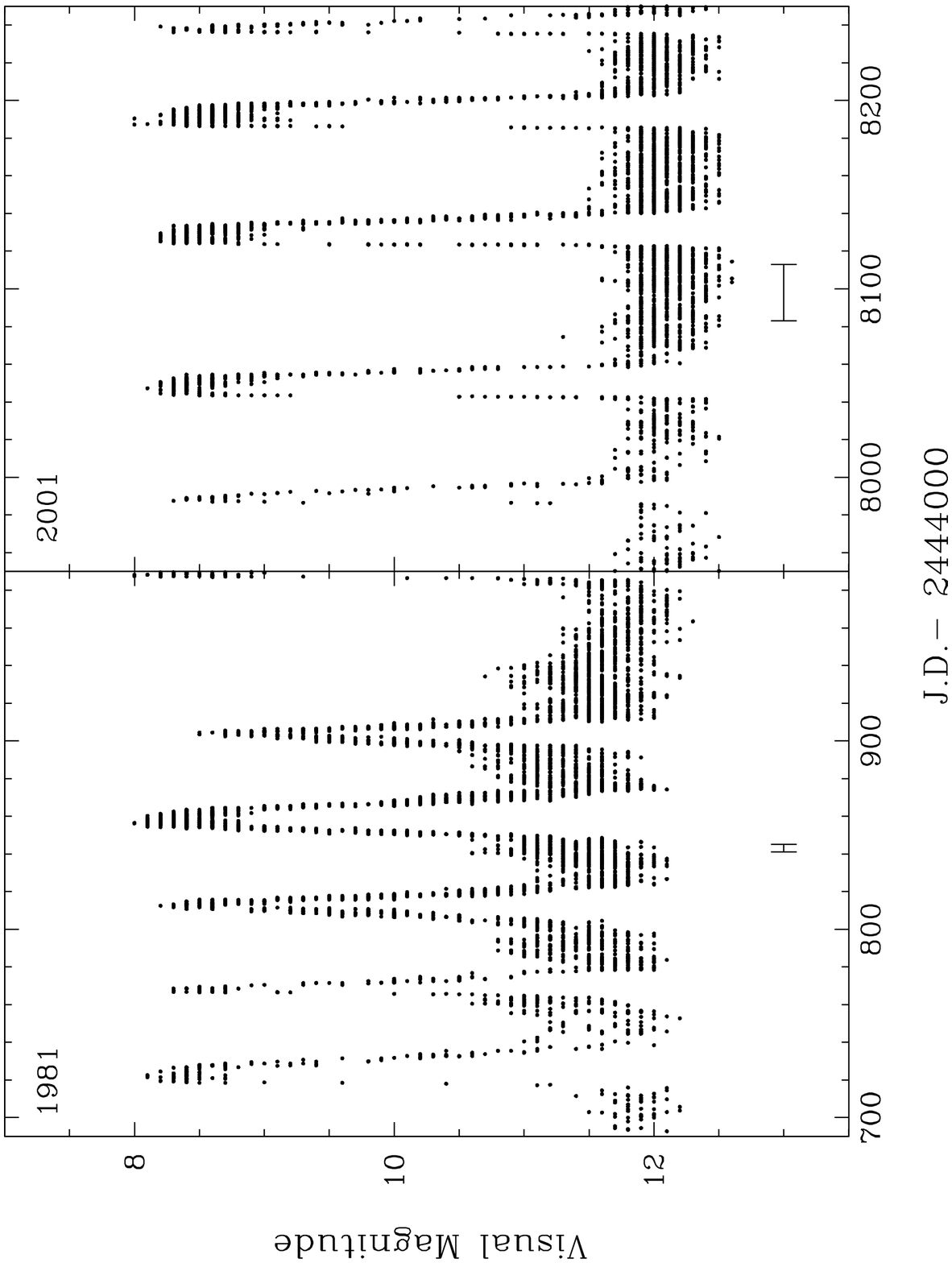]
{The eruption light curve of SS~Cyg in 1981 and 2001 from
the American Association of Variable Star Observers.
The horizontal bar plotted below the light curve in 1981
marks the dates when \citet{hessman84} obtained their 
spectrograms of SS~Cyg; the horizontal bar in 2001 shows
when we obtained our spectrograms.
Both sets of data were obtained when SS~Cyg was in its
quiescent state, but SS~Cyg was $\sim \! 0.6$~mag
fainter during quiescence in 2001 than it was in 1984.
\label{EruptionLC-fig}
}

%  FIGURE 2
\figcaption[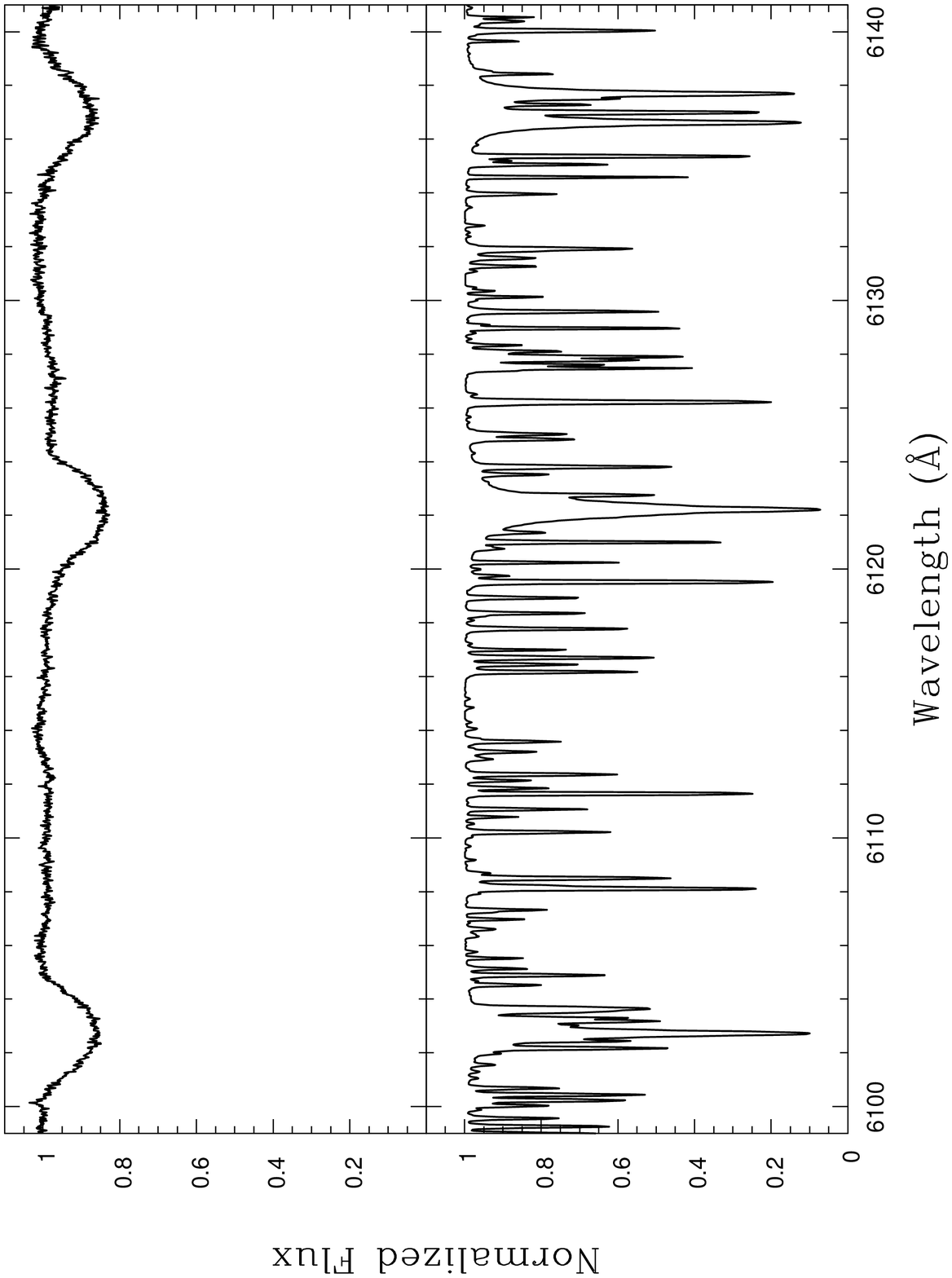]
{The upper panel shows a portion of the average of all our
spectrograms of SS~Cyg.  
Before the spectrograms were averaged their orbital velocities 
were removed by shifting the spectrograms to SS~Cyg's systemic 
velocity.
The lower panel shows a synthetic spectrum for a single, non-rotating
star with a K4-5~V spectral type for comparison.
The extra width of the absorption lines in SS~Cyg is due to a
combination of rotational broadening ($\sim \! 89\ {\rm km\ s}^{-1}$),
orbital smearing during the exposures ($<26\ {\rm km\ s}^{-1}$),
and instrumental resolution of the spectrograph
($8.2\ {\rm km\ s}^{-1}$).
\label{Spectrogram-fig}
}

%  FIGURE 3
\figcaption[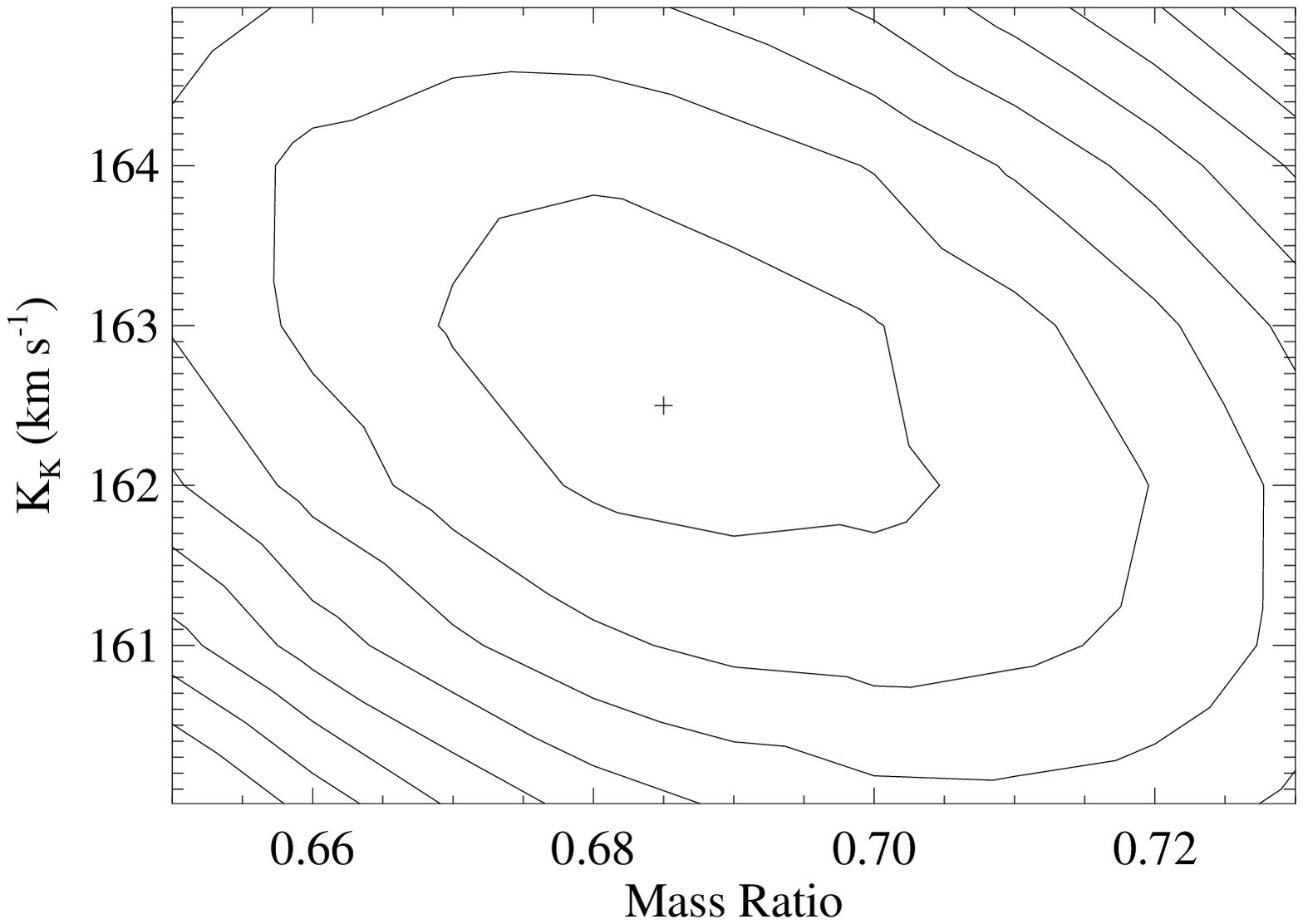]
{A contour plot of the $\chi^2$ of the fits of the LinBrod
synthetic spectra to the observed spectrum of SS~Cyg
as a function of $K_K$ and $q$.
The contours of constant $\chi^2$ are evenly spaced
multiples of the minimum $\chi^2$ beginning at 1.0001
and separated by 0.0002.
The best fit values are marked by the cross at
$q = 0.685$ and $K_K = 162.5\ \textrm{km s${-1}$}$.
\label{KqChi-fig}
}

%  FIGURE 4
\figcaption[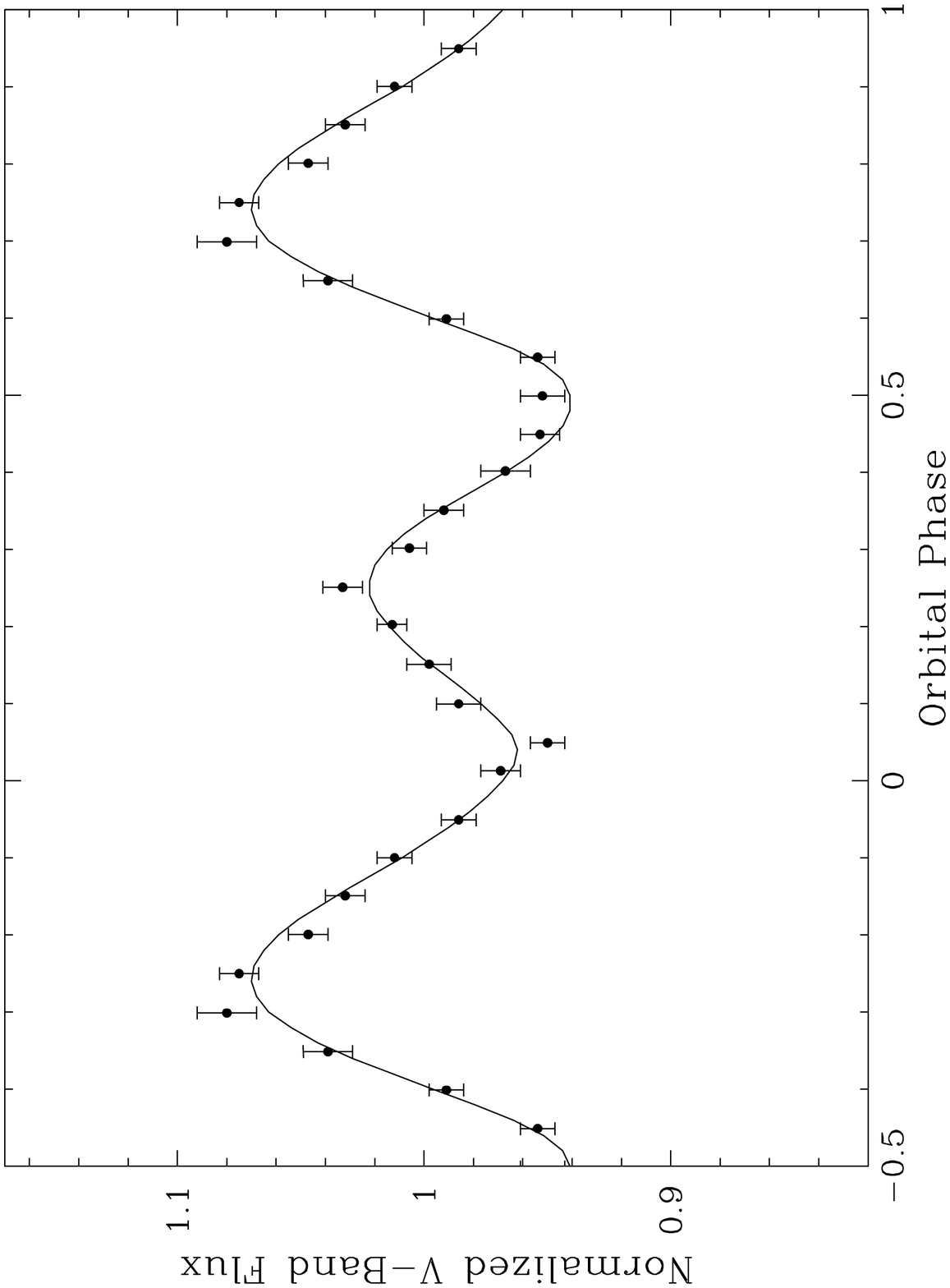]
{The dots and error bars are measurements of the V-band orbital 
light curve of SS~Cyg from \citet{voloshina00}.
Orbital phase $\phi = 0$ corresponds to spectroscopic 
conjunction with the white dwarf behind the K star.
The basic double-humped variation is produced by
ellipsoidal variations of the K star.
The extra amplitude and the asymmetry of the hump at 
$\phi = 0.75$ is produced by a bright spot 
on the outer edge  of the disk.
The solid line is a fitted light curve for an orbital
inclination $i = 52^\circ$ and 56\% of the light at  
$\phi = 0.25$ coming from the accretion disk.
As noted in the text and shown in Figure~\ref{inclination-fig},
there is a strong positive correlation between the adopted
third-light fraction and the best fitting orbital inclination.
\label{LCfit-fig}
}

% FIGURE 5
\figcaption[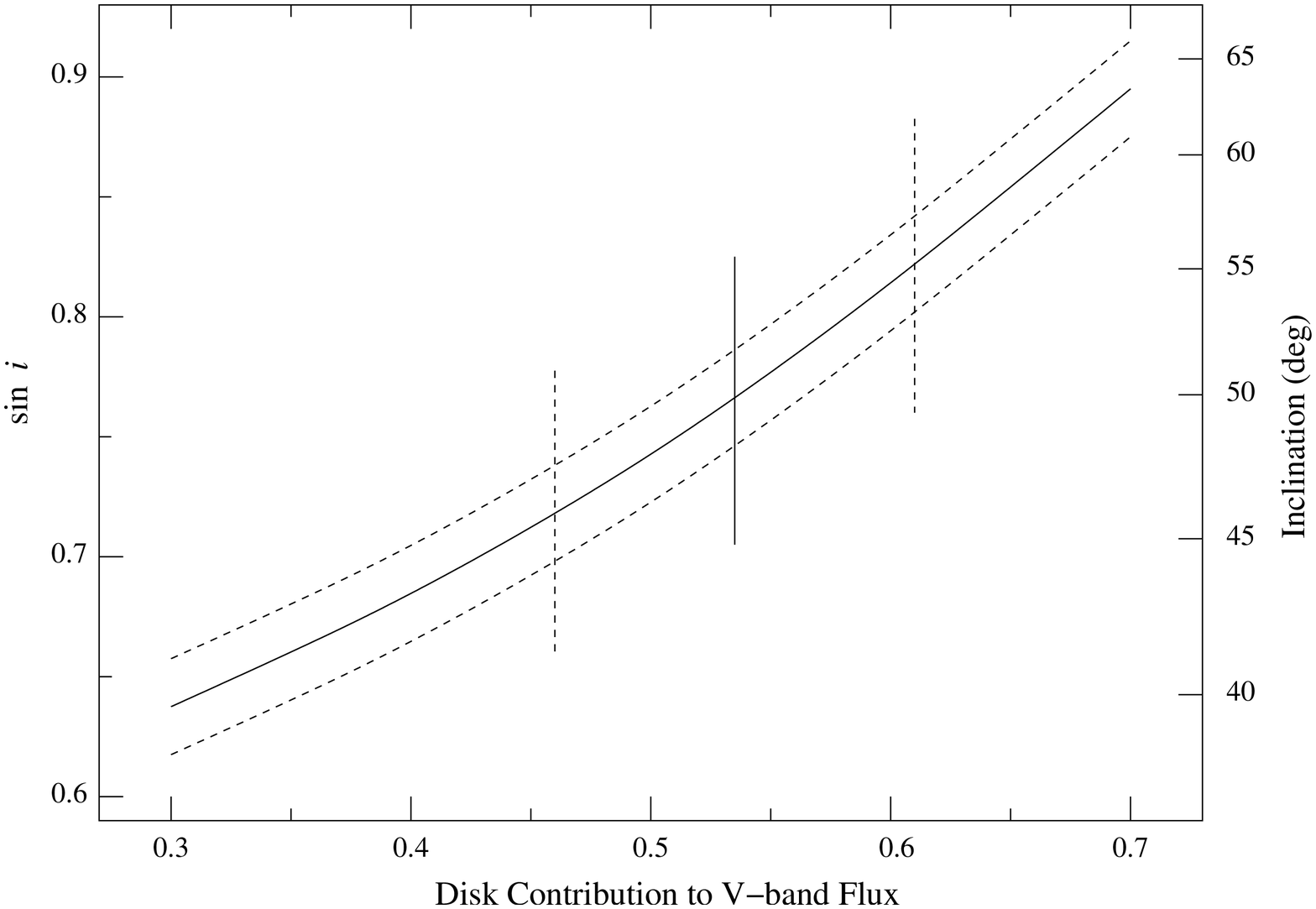]
{
The correlation between orbital inclination and the fraction
$f$ of the V-band flux contributed by the accretion disk 
and white dwarf.
The best fitting inclinations for any adopted 
fraction are shown by the solid line extending from the lower-left
to the upper right in the diagram.
The parallel dashed lines show one standard deviation confidence
limits for the fits.
The vertical solid corresponds to $f = 0.535$, derived from the 
dilution of the absorption lines in the spectrum of the K star;
and the two vertical dashed lines correspond to 
$f = 0.535 \pm 0.075$, the confidence limits on f.
The derived inclination lies in the interval 
$45^\circ \le i \le 56^\circ$.
\label{inclination-fig}
}

% FIGURE 1
\begin{figure}
\figurenum{\ref{EruptionLC-fig}}
\plotone{f1.eps}
\end{figure}

% FIGURE 2
\begin{figure}
\figurenum{\ref{Spectrogram-fig}}
\plotone{f2.eps}
\end{figure}

% FIGURE 3
\begin{figure}
\figurenum{\ref{KqChi-fig}}
\plotone{f3.eps}
\end{figure}

% FIGURE 4
\begin{figure}
\figurenum{\ref{LCfit-fig}}
\plotone{f4.eps}
\end{figure}

% FIGURE 5
\begin{figure}
\figurenum{\ref{inclination-fig}}
\plotone{f5.eps}
\end{figure}

\end{document}